\documentclass[a4paper,11pt]{article}

\usepackage[utf8]{inputenc}
\usepackage[T1]{fontenc}
\usepackage{graphicx} 
\usepackage[english]{babel}
\usepackage{amsfonts} 
\usepackage{amsmath}  
\usepackage{verse}
\usepackage{appendix}
\usepackage{enumitem}

\begin{document}
  \title{\textbf{The influence of the migration network topology on the stability of a small food web}}%
  \author{JONAS RICHHARDT$^*$, SEBASTIAN PLITZKO, \\
 \textit{Institute of condensed matter physics,}\\ \textit{Darmstadt University of Technology, 64289 Darmstadt, Germany}
\and
FLORIAN SCHWARZM\"ULLER \\
 \textit{J.F. Blumenbach Institute of Zoology and Anthropology,}\\ \textit{Georg-August-University
G\"ottingen, 37073 G\"ottingen, Germany}\\
AND \\ BARBARA DROSSEL \\
\textit{Institute of condensed matter physics,}\\ \textit{Darmstadt University of Technology, 64289 Darmstadt, Germany}
\\ $^*$Corresponding author: richhardt@fkp.tu-darmstadt.de
}
\maketitle

\begin{abstract}
The stability of ecosystems as well as the relation between topology and dynamics on multilayer networks are important questions that are usually discussed in separate communities. Here, we combine these two topics by investigating the influence of the topology of the migration network on the stability of a four-species foodweb module on six patches. The parameters are chosen such that the dynamics on an isolated patch have a periodic attractor with all four species present as well as an attractor where the prey that is preferred by the top predator dies out. The stability measure used here is robustness, which is the average proportion of surviving species in the system, and which shows a complex dependence on the migration rate. We use principal component analysis to quantify the migration network structure in terms of the most relevant network measures, and we evaluate correlations between these measures and characteristics of the robustness curves. Our most important findings are that higher connectivity of the migration network leads to a larger maximum robustness, that a broad distribution of connectivities favors extinction of the preferred prey at intermediate migration rates, and that migration topologies with a larger betweenness centrality are more prone to extinction of the preferred prey at the onset of synchronization. Our study thus demonstrates a strong correlation between dynamical robustness and spatial topology and can serve as an example for similar studies in other types of multilayer networks.
 
\end{abstract}

\section{Introduction}

The relation between the topology of networks and the dynamics on these networks is the subject of intense research \cite{barrat2008dynamical, hutt2014perspective} due to its relevance to a wide range of fields, such as disease spreading, neural network firing patterns, electric power failure cascades, and species survival in ecological systems.  In the context of ecology, understanding the effect of space on the stability of food webs is of prime importance for ecosystem conservation and management. Food webs are directed networks, where the connections between species represent feeding interactions. The stability of food webs has been investigated for a long time since the seminal study by May \cite{may1971}, who found by a linear stability analysis based on random connectivity matrices that more complexity leads to less stability. Since then, the influence of features such as realistic foodweb topologies \cite{yodzis}, weak links  \cite{mccann}, or allometric scaling \cite{brose} on stability was investigated. More recently, the focus has turned on the influence of the spatial structure on the dynamics of foodwebs, which was often neglected in earlier foodweb studies, but is very relevant for predator-prey interactions \cite{abrams}. Historically random migration, which is essentially a random walk or diffusion process, was the first attempt to model migration \cite{skellam}, and it is still widely used in modelling efforts, despite the fact that adaptive migration appears more realistic \cite{bowler}. Many results for food-web modules consisting of three species are compiled in a recent review by Amarasekare \cite{amarasekare}, where the effect of food web complexity and type of movement on synchrony and stability is summarized and explained by underlying mechanisms. 

This paper is concerned with systems of four species, which are connected to form the diamond motif. This is a common food web motif, where a top species feeds on two intermediate species, which in turn share a resource. Most consumers in nature share resources and predators with other consumers, making this module a useful simplification of many natural food webs.  The importance of such motifs for biological networks in general is
emphasized by Milo et al \cite{milo}, showing that the diamond motif is much more frequent in biological networks than in randomly generated networks that have the same number of nodes and connections.  In contrast to studies of this motif on large regular spatial networks \cite{maser07,gouhier10}, we consider systems of only a few patches in order to explore the relation between the topology of the migration network and the dynamical stability, using two types of random migration. In a recent work by Ristl et al. \cite{plitzko}, the robustness curve, which is obtained when evaluating robustness as function of migration rate, was evaluated for systems of 2 to 10 patches for selected topologies. This robustness curve has a surprisingly complex, nonmonotonous shape, which lends itself to evaluating its characteristics as function of the network topology. We therefore explore the correlation of significant features of the robustness curves with different topological measures of the six-patch network, considering all 112 different topologies. 

In the next section, we explain in detail the model and simulation setup used for our study. In section \ref{measures}, the quantitative measures of the robustness curves and the migration topologies that are most suitable for our correlation study are determined. The results of this study are presented and explained in section \ref{results}, and section \ref{conclusion} places these findings in a wider context.

\newpage

\section{Model}
\label{model}

\begin{figure}
\centering
\includegraphics[width=2.5cm]{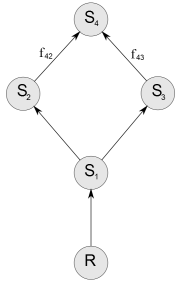}
\caption{Diamond food web: The predator prefers S$_2$ which is implemented by setting $f_{42} = 0.6$ and $f_{43} = 0.4$. The arrows denote the flow of biomass.}
\label{diamond}
\end{figure}

We focus on the diamond food-web configuration shown in Figure \ref{diamond}, in which two consumers (S$_2$ and S$_3$) compete for their common resource (S$_1$) and are consumed by a generalist predator (S$_4$) that has a preference for S$_2$. This module was recently investigated on a system of two to ten patches for selected patch topologies \cite{plitzko}, and we will use exactly the same dynamical model with the same parameters.
The consumers and the predator exhibit a type II functional response, and the resource (S$_1$) undergoes logistic growth. This logistic growth is implemented in the computer simulations by providing a constant external resource (R) as food source for S$_1$. In this way, the population dynamics equation of S$_1$ has
the same form as that of the other three species. 

This diamond food web model is placed on each of the six patches, which are coupled via migration. Migration is modelled as a diffusion process, with the rate of emigration being proportional to the population size of a patch.

The time evolution of the biomass density $B_{i,k}$ of species $i$ in patch $k$ is described by the equation
\begin{align}
\dot{B}_{i,k}(t) = \text{ } &\lambda \sum_{j \in R_i} g_{ij,k}(t)B_{i,k}(t) -\alpha B_{i,k}(t) - \beta B^2_{i,k}(t) - \sum_{j \in C_i} g_{ji,k}(t)B_{j,k}(t) \notag\\
&+ \sum_{l \neq k} d(\xi_{lk}B_{i,l}(t)- \xi_{kl}B_{i,k}(t)),
\label{popdyn}
\end{align}
with a Holling type II functional response
\begin{align}
g_{ij,k} = \frac{af_{ij}B_{j,k}}{1+\sum_{n \in R_i} h a f_{in} B_{n,k}}.
\end{align}

Here, $\lambda$ is the assimilation efficiency, $\alpha$ is the respiration rate, $\beta$ is the intraspecific competition strength, $h$ is the handling time and $a$ is the attack rate. $R_i$ and $C_i$ denote the set of prey and the set of predators of the species with index $i$. $f_{ij}$ denotes the foraging efforts. Only $f_{42}$ and $f_{43}$ differ from 1. The external resource R is included in these equation as species 0 with a constant biomass $B_{0,k}=500$.
The parameter values are $\alpha = 0.3$, $\beta = 0.5$, $\lambda = 0.65$, $h = 0.35$,  as in \cite{plitzko} and \cite{heckmann}. The selection of these parameters is due to empirical bio-energetic considerations \cite{brose}. The foraging efforts are set to $f_{24} = 0.6$ and $f_{34} = 0.4$, so that there are two attractors of the dynamics. The first attractor is a limit circle, where the biomass of all four species oscillates. The second attractor is a fixed point where the species S$_2$ becomes extinct, leaving a tritrophic food chain. Attractors where more species die out are so rare ($\leq 1$ out of 1000) that they can be neglected.  

The last term in equation (\ref{popdyn}) describes migration between patches. The product $d\xi_{ij}$ is the migration rate from patch $i$ to patch $j$. $d$ denotes the migration strength and the parameters $\xi_{ij}$ denote the link strengths. We investigated all 112 possible topologies of a six-patch network. 112 is the number of non-isomorphic connected undirected graphs witch 6 vertices. 
The topologies are numbered from $T=0$ (star) to $T=111$ (fully connected) roughly in the order of increasing degree, as they are generated by the program nauty \cite{McKay201494}. Some of these topologies are given with their numbers in Figure \ref{robcases}. A few other topologies, which are prominent in Figures \ref{lwpc1allinone} to \ref{pwpc2allinone} below, are a star with one additional connection ($T=2$), chain ($T=23$), ring ($T=48$), a triangle with two tails of length 1 and 2 attached to the same node ($T=15$), a quadrangle with two spines attached to neighboring nodes ($T=17$), a triangle with a spine attached to each node ($T=19$), a star with two additional connections between two distinct pairs of satellite nodes ($T=16$) , a star with three  additional pairwise connections within a group of three satellite nodes ($T=33$), a ``kite'' with a fully connected set of four nodes and a tail of two nodes ($T=78$), a ``kite'' where one of the inner diagonal connections is removed ($T=29$ and $T=75$), the topologies where each node is connected to 3 other nodes ($T=70$ and $T=92$), and the topology where each node is connected to 4 other nodes ($T=108$).

\begin{figure}[htp]
\centering
\includegraphics[width=12cm]{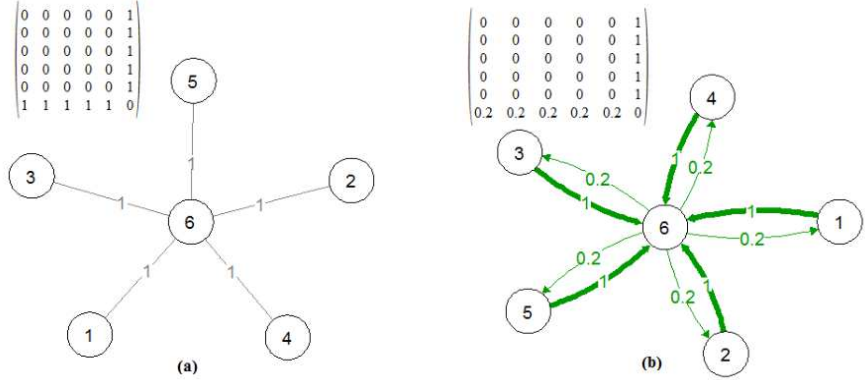}
\caption{Star migration network: The central patch (6) is surrounded by five satellite patches.  \textbf{(a)} Linkwise migration network with the corresponding adjacency matrix of the link strengths $\xi_{ij}$.
\textbf{(b)} Patchwise migration network with the corresponding adjacency matrix of the link strengths $\xi_{ij}$.}
\label{lwpwstar}
\end{figure}

We distinguish between two ways of modelling the link strengths. For ``linkwise migration'', all nonvanishing link strengths are unity. For ``patchwise migration'', we set $\xi_{ij}=1/z_i$, with $z_i$ being the degree of patch $i$. This means that each patch has the same total migration rate. Figure \ref{lwpwstar} illustrates the difference between patchwise and linkwise migration for the special case of the star topology. In the context of graph theory a linkwise migration network corresponds to an undirected graph, while a patchwise migration network corresponds to a directed weighted graph, since the link strength between a pair of patches is sensitive to the orientation. In graph theory, the parameters $\xi_{ij}$ are the entries of the adjacency matrix.

For a particular migration topology the migration strength $d$ is the only parameter that is varied. We evaluated the stability of the system in terms of the robustness, i. e. of the ratio of the number of surviving species after the simulation and the number of initial species. A robustness value of 1 means that the dynamics settle for all initial conditions on the oscillating attractor with four species. A robustness values of 0.75 means that for all initial conditions species $S_2$ dies out, leading to the tritrophic food chain. Robustness values between 0.75 and 1 indicate that each of these attractors is reached for part of the simulation runs. We performed simulations for 6000 time units. We set the extinction threshold, i. e. the biomass density below which a species is considered to be extinct to $10^{-6}$. We carried out at least 1000 runs for each type of migration and for each value of the migration strength, with the initial biomasses chosen randomly from the interval $\left[0,3\right]$.

\section{Relevant quantitative measures}
\label{measures}

In order to investigate the correlation between robustness and patch topology, we identify in the following useful quantitative measures for both.

\subsection{Characteristic features of the robustness curves}

\begin{figure}[htp]
\centering
\includegraphics[width=12cm]{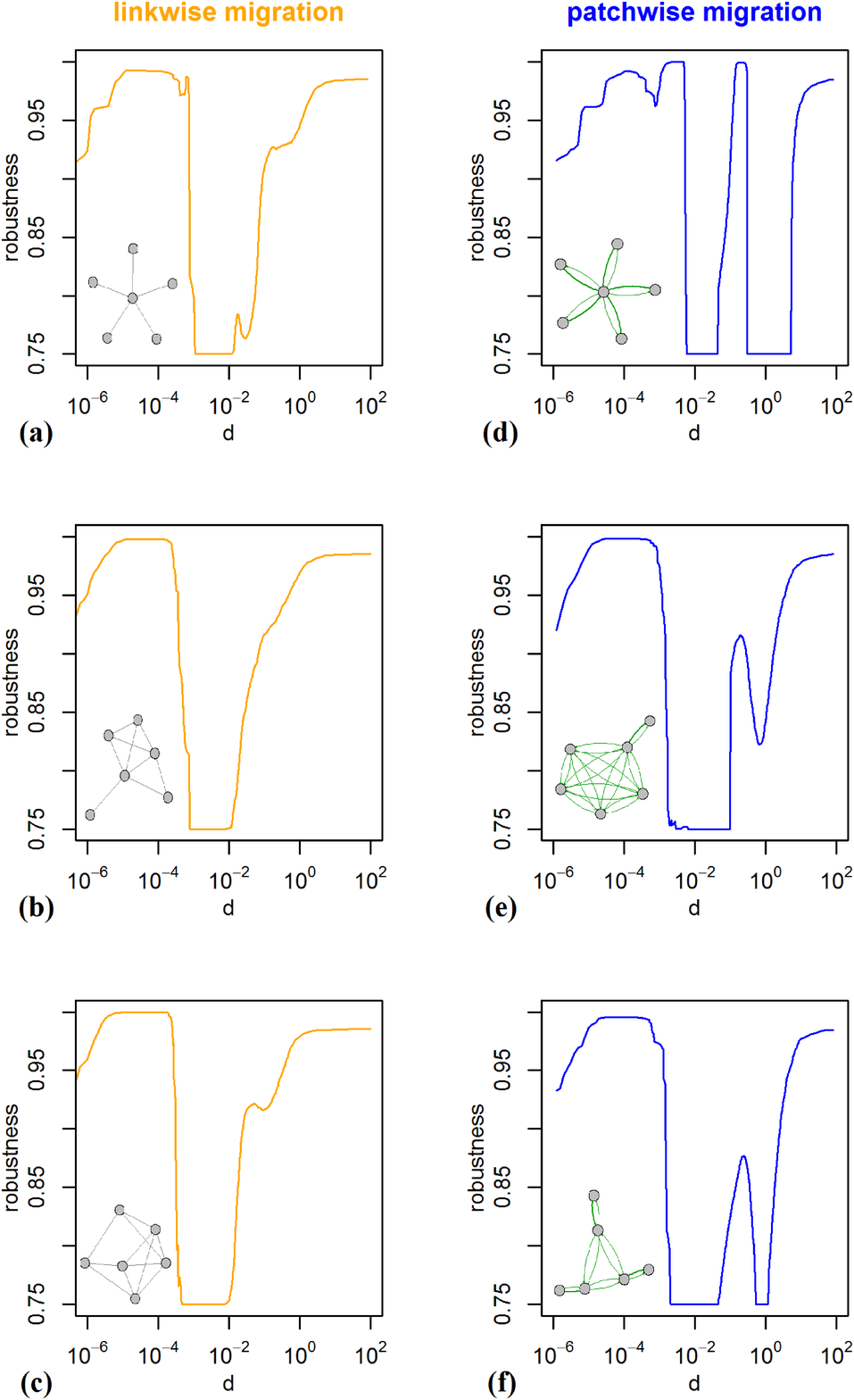}
\caption{Robustness curves for selected topologies. \textbf{(a)} star topology ($T = 0$) with linkwise migration. \textbf{(b)} topology $T = 42$ with linkwise migration. \textbf{(c)} topology $T = 106$ with linkwise migration. \textbf{(d)} star topology ($T = 0$) with patchwise migration. \textbf{(e)} topology $T = 101$ with patchwise migration. \textbf{(f)} topology $T = 19$ with patchwise migration.   }
\label{robcases}
\end{figure}

As discussed in detail in recent work by Ristl et al. \cite{plitzko},
 the robustness shows a complex dependence on the migration strength $d$. Figure \ref{robcases} gives an impression of the variation between the robustness curves for selected patch topologies. These curves share several features: For $d$ below approximately $10^{-6}$, they approach the robustness value $\simeq 0.903$ of an isolated patch. With increasing $d$, robustness increases due to the rescue effect, which prevents extinction of S$_2$ by immigration from neighboring patches. At intermediate migration strength, when the time scale of migration becomes comparable to that of in-patch population dynamics, dynamics become more complex, and there is an interval of $d$ values for which species S$_2$ dies out for all initial conditions. This manifests itself in a plateau with the robustness value 0.75 found typically around $d$ values of $10^{-2}$. Depending on the type of migration and the topology, a second such plateau occurs at larger $d$ values, or at least a local minimum with reduced robustness. For linkwise migration, there is never a second plateau. For patchwise migration and the star topology, the robustness peak between the two plateaus is highest and reaches the value 1.  When migration is much faster than population dynamics, i.e. $d > 10$, the robustness curves approach an asymptotic value, which is close to 0.985. For linkwise migration, very fast migration then makes the population sizes on all patches identical. The patches are therefore synchronized and show the same attractors as an isolated patch. The asymptotic robustness value is nevertheless different from that of an isolated patch since the initial population sizes are now given by the average of six (instead of one) random variables taken from constant distributions. For patchwise migration, very fast migration distributes the populations such that there is no net migration through any connection. This means that the population sizes divided by the degree of the patch are identical on all patches. For topologies with the same degree for each patch, the robustness curves for patchwise migration become identical to those for linkwise migration when $d$ is divided by the degree of the patches.

We chose the following features of the robustness curves for our evaluation of the correlation with features and the patch topology:
\begin{enumerate}
\item "First 0.75": Migration strength at which the robustness reaches 0.75 for the first time.
\item  "End of first plateau": Migration strength at the end of the first plateau of the robustness curve.
\item "Final position": Migration strength at which the robustness curve reaches 0.99\% of the asymptotic value obtained for $d = 100$.
\item   "Global maximum value": Value of the global maximum of the robustness curve, which is usually found for values of $d$ close to $10^{-4}$.
\item $\Lambda$: The difference between the local maximum and the local minimum robustness value at the right-hand side of the plateau for all topologies which show a second minimum of the robustness curve. This quantity is only evaluated for patchwise migration.
\end{enumerate}
For the first three of these quantities, we also define the ratio between their values for patchwise and linkwise migration, normalized by the mean degree,
\begin{equation}
\Delta(x,T) = \frac{x_{pw}}{x_{lw}} \frac{1}{\text{mean degree}(T)}. \label{Delta}
\end{equation}
Here, $x$ denotes the quantity, and $T$ denotes the topology.

\subsection{Characteristic features of the migration network}

In order to quantify the migration network structure, we borrowed concepts from graph theory. In graph theory, our patches correspond to vertices, and our connections to edges. We evaluated the following 12 network features:
\begin{enumerate}
\item The mean degree of all vertices, i.e., the mean number of outgoing edges (``mean.dg'')
\item The maximum degree of all vertices (``max.dg'')
\item The standard deviation of the degree (``sd.dg'')
\item The average path length: the length of the shortest path between two vertices, averaged over all pairs of vertices and weighted by the sum of the link strengths that are passed (``av.path.length'')
\item The maximum path length (``max.path.length'')
\item The standard deviation of the path length (``sd.path.length'')
\item The modularity 
\begin{equation}
Q = \frac{1}{2m}\sum_{i,j}\left( \xi_{ij} - \frac{k_ik_j}{2m} \right)\delta_{c_ic_j}\, ,
\end{equation}
 with $m$ being the total number of edges, $k_i$  being the sum of the edge weights  of  all adjacent edges for vertex $i$, and $c_i$  denoting the component to which vertex $i$ belongs. Components were determined using the walktrap community finding algorithm based on a random walk as implemented in the igraph package \cite{igraph}. The idea of the algorithm is that short random walks are likely to stay in the same community. 
The modularity is a measure for the existence of connected subgraphs and is largest (about 0.38) for the graph that contains two triangles connected by one edge, and zero for the star and the fully connected graph.
\item The transitivity, which is the probability that the neighbours of a vertex in the network are connected themselves.
\end{enumerate}

Additionally, we measured centrality of the graphs based on vertex properties. The single graph-level score of a centrality measure C is derived by the formula $C = \sum_v \text{ max}_w (C_w) - C_v$, where $C_v$ is the vertex-level centrality score. This leads to a high value of graph-level centrality when vertices are very different in this property. Besides $C$ is normalized to one. We considered the following centrality scores:

\begin{enumerate}[resume]
\item  The degree centrality, where the vertex property is simply the degree. 
\item The betweenness centrality, where $C_v$ is given by the number of shortest paths from all vertices to all others that pass through vertex $v$.
\item The closeness centrality, where $C_v$ being the inverse of the sum of the closest distances from vertex $v$ to all other vertices.
\item The eigenvector centrality, where $C_v$ of vertex $v$ is the $v$th entry of the eigenvector to the largest eigenvalue of the adjacency matrix. It tends to be large if the vertex is part of a clique.
\end{enumerate}

We evaluated these 12 quantities for all 112 linkwise and all 112 patchwise topologies using the igraph package \cite{igraph}. Note that eigenvector centrality is not defined for patchwise migration, since the corresponding adjacency matrix is not symmetric. The 12 quantities are not independent from each other. Our evaluation showed that there is a strong positive correlation between the distinct centralities. There is also a strong positive correlation between the three properties of the path length distribution. In addition there is a strong negative correlation between the mean degree and the average path length. 
In order to handle this interrelated data set we used principal component analysis (pca). The central idea of pca is the reduction of dimensionality of a correlated data set by transforming to a new set of variables - the principal components \cite{jolliffe}. As required for pca, we normalised the data sets so that they share a vanishing arithmetic mean of zero and a standard deviation of unity. We found that in the case of linkwise migration already 2 principal components (pc) suffice to describe $84.3\%$ of the variance of the original twelve-dimensional data set. The first pc contributes already nearly $50\%$ of the total variance. In the case of patchwise migration, the first two pc explain $85.6\%$ of the original variance.

\begin{figure}[htp]
\centering
\includegraphics[width=9cm]{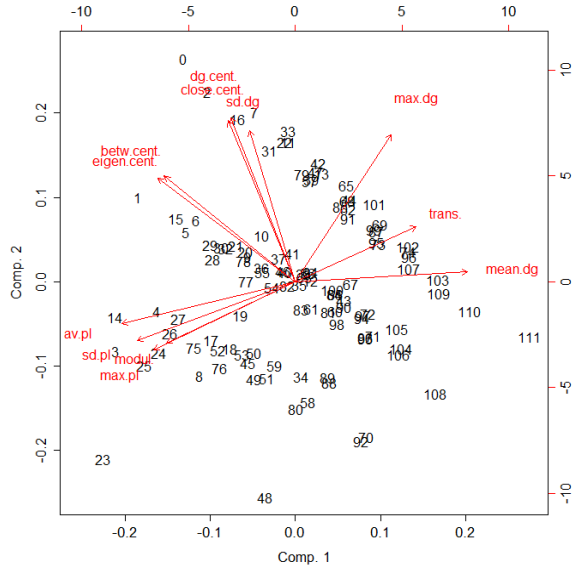}\\
\includegraphics[width=9cm]{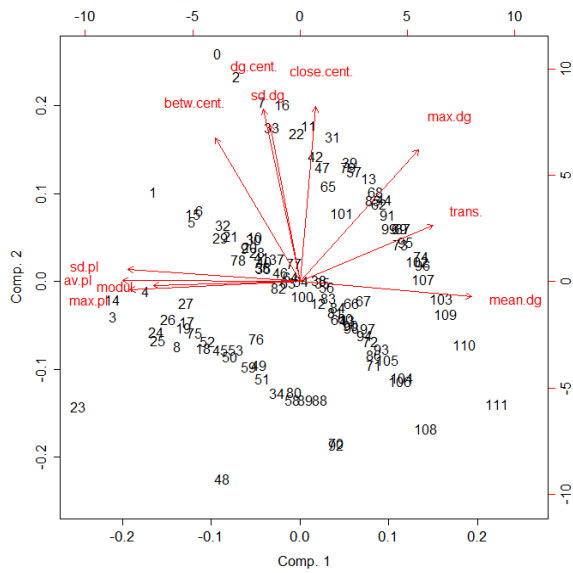}
\caption{Biplot of the first two pc for linkwise (top) and patchwise (bottom) migration.}
\label{biplots}
\end{figure}

Figures \ref{biplots} shows the planes that are spanned by the first two pc for linkwise and patchwise migration. One can see that the first pc is almost parallel to the mean degree and antiparallel to the mean path length in both cases. The second pc is related to degree centrality, closeness centrality and the standard deviation of the degree distribution. It is large when the degrees of the vertices differ a lot. The figures show the above-mentioned strong correlations between different measures. They also show that the networks are divided into 3 groups that are arranged in stripes. These groups differ by the maximum degree, which is 5,4,3 (from top to bottom). The two networks with maximum degree 2 (chain, $T=23$, and ring, $T=48$) can be found below the three stripes.

\section{Correlation between robustness and patch topology}
\label{results}

Figures \ref{lwpc1allinone} to \ref{pwpc2allinone} show the first four measures of the robustness curves as function of the two pc for all 112 topologies. We describe and discuss in the following these results separately for linkwise and patchwise migration, and then we compare the two. 

\subsection{Linkwise migration}
\begin{figure}[htp]
\centering
\includegraphics[width=13cm]{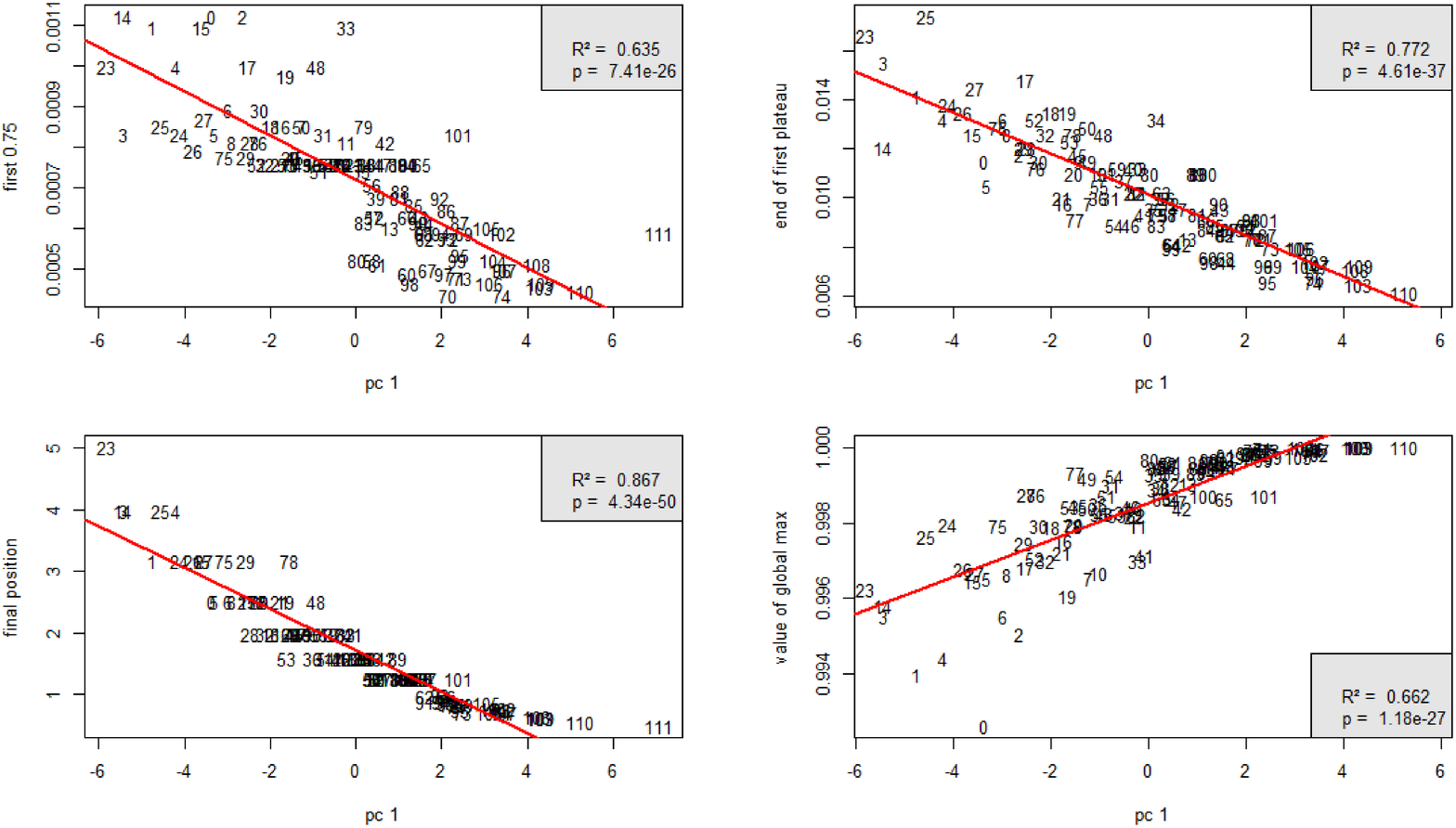}
\caption{Characteristics of the robustness curves with linkwise migration plotted over pc 1.}
\label{lwpc1allinone}

\centering
\includegraphics[width=13cm]{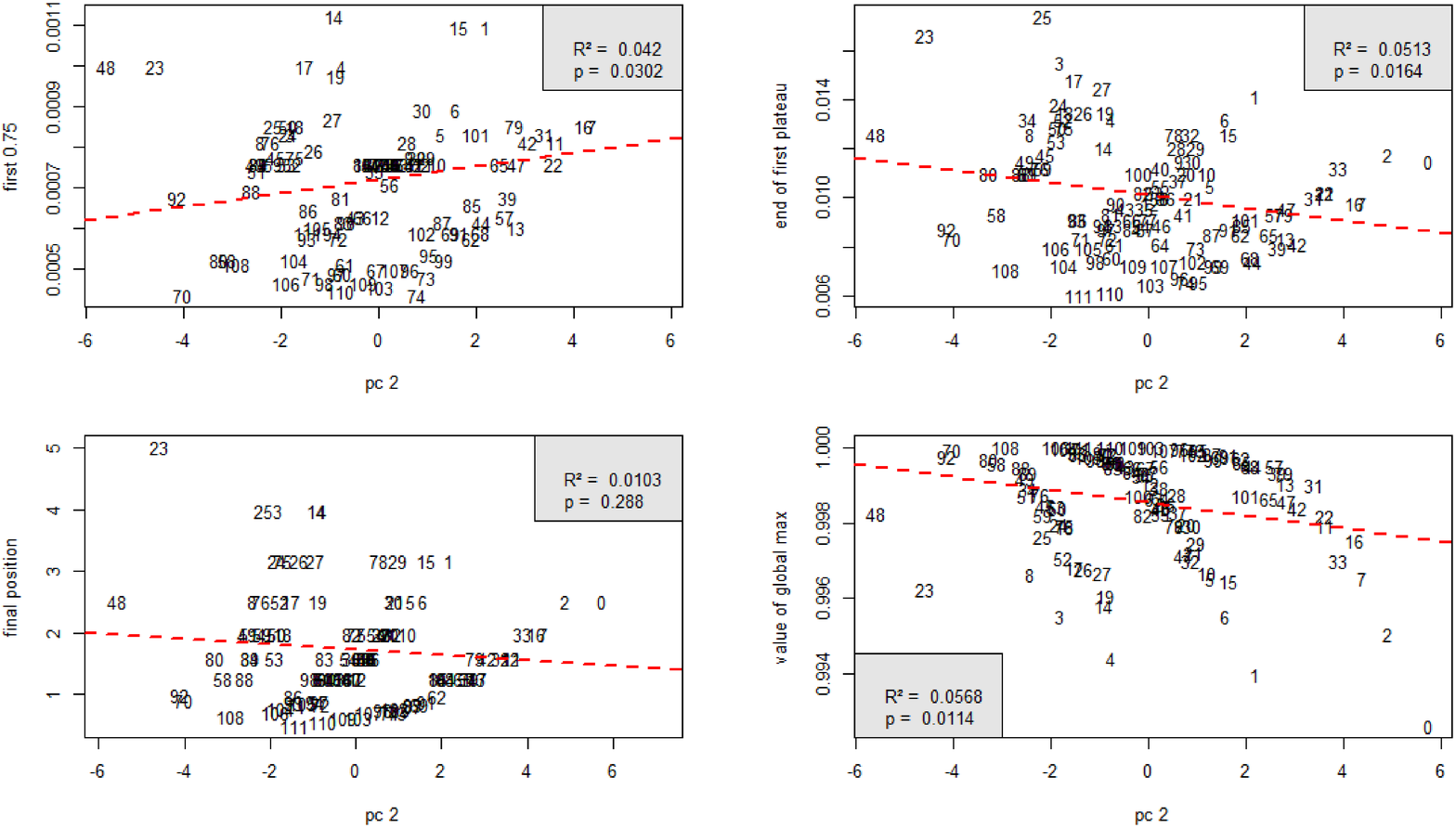}
\caption{Characteristics of the robustness curves with linkwise migration plotted over pc 2.}
\label{lwpc2allinone}
\end{figure}

Both the migration strength  at which the robustness reaches $0.75$ for the first time ("first 0.75") and the position where the robustness leaves the plateau ("end of first plateau") decreases with the first pc (see Figure \ref{lwpc1allinone}). The same holds for the migration strength where robustness reaches the asymptotic value "final position". This means that these three quantities decrease with increasing mean degree, since the first pc is essentially the mean degree. We ascribe this to the fact that a larger mean degree means more migration into a patch and out of a patch for a given value of the migration strength $d$.

The value of the global maximum shows the opposite trend and increases with the mean degree. Since the global maximum usually occurs in the $d$ range where the rescue effect is operating, this result is plausible: when a patch, where initial population sizes drive species S$_2$ towards extinction, is connected to more other patches, there is a  larger change of rescuing species S$_2$ by immigration from a neighboring patch. 

In contrast, there is no systematic trend in these four measures when they are plotted against pc2. This is not surprising, since pc 1 describes the data sets rather well, and principal components are uncorrelated by construction. All four plots in Figure \ref{lwpc2allinone} show the three stripes that correspond to different maximum degree, and within each of these stripes there is a clear trend of the data, but this is again due to the change of pc1 as one moves along a stripe. 

\subsection{Patchwise migration}
\begin{figure}[htp]
\centering
\includegraphics[width=13cm]{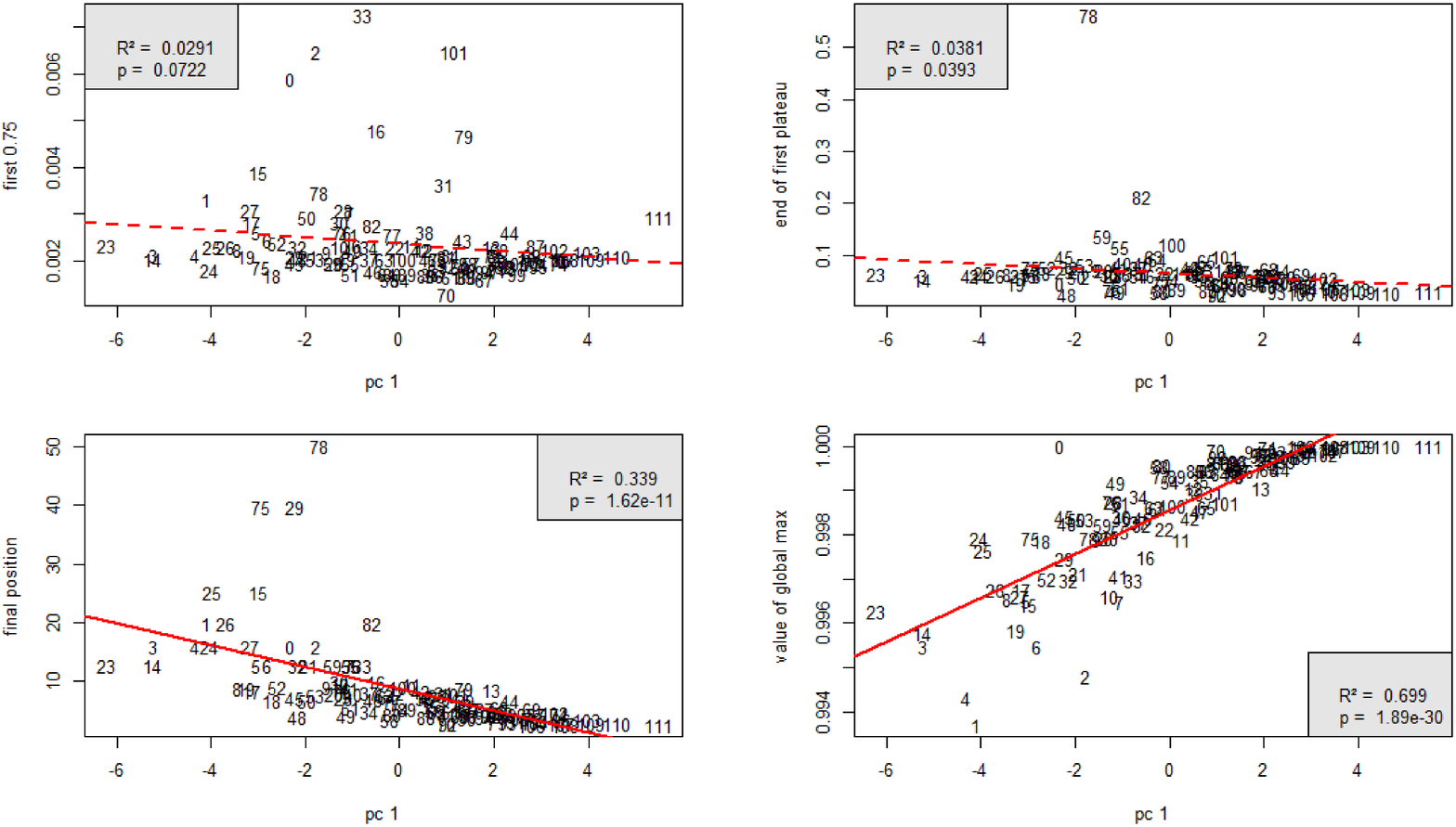}
\caption{Characteristics of the robustness curves with patchwise migration plotted over pc 1.}
\label{pwpc1allinone}

\centering
\includegraphics[width=13cm]{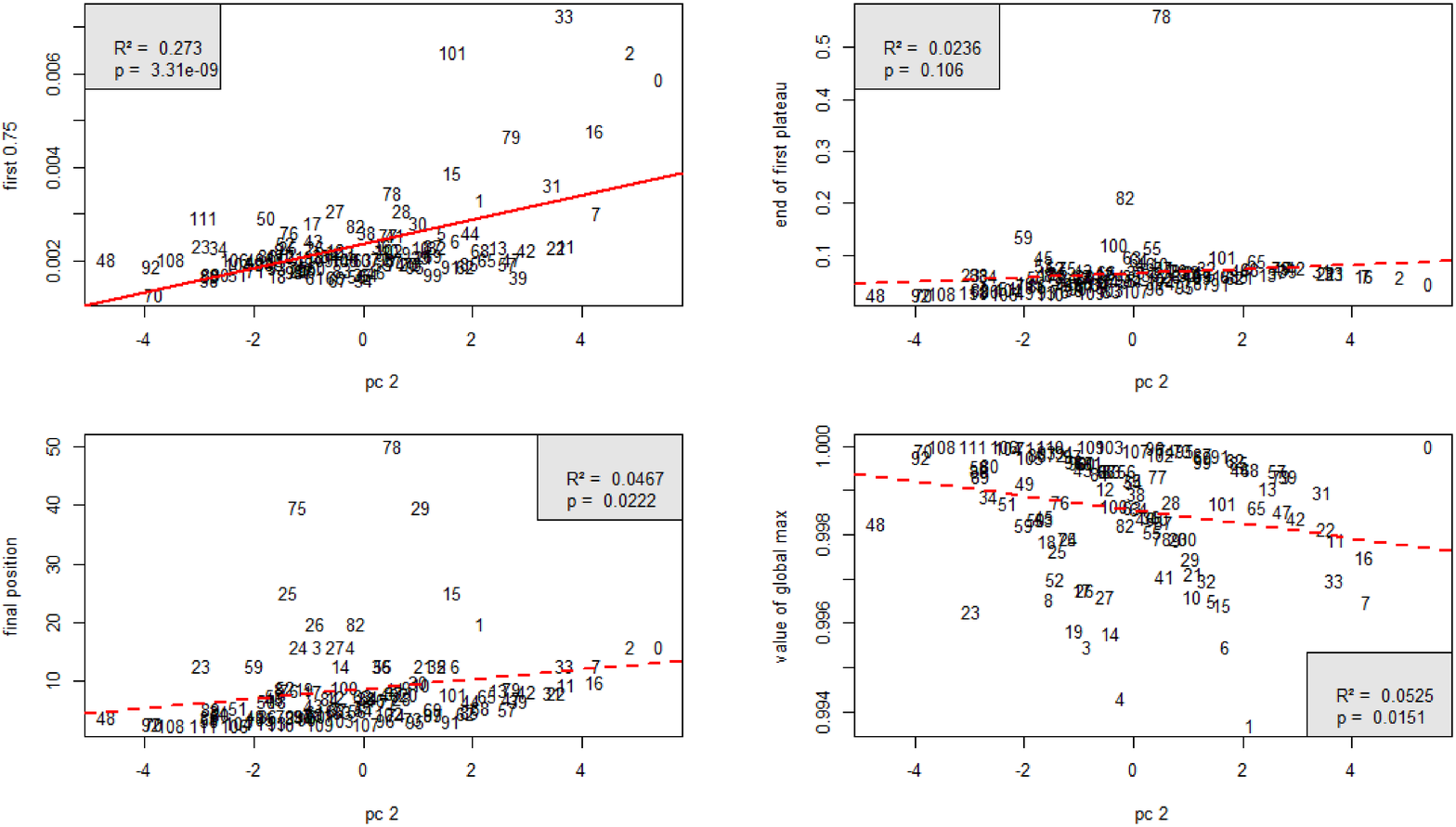}
\caption{Characteristics of the robustness curves with patchwise migration plotted over pc 2.}
\label{pwpc2allinone}
\end{figure}

In contrast to linkwise migration, there appears to be no clear correlation between "fist 0.75" and pc 1 (see Figure \ref{pwpc1allinone}). The plotted straight line explains only about $3\%$ of the data variance as indicated by the value of $R^2$, and the $p$ value (about $7\%$) is quite high. The $p$ value corresponds to the probability that an event that is at least as extreme as the present did arise randomly if the actual relation were a straight line with slope 0. Similarly, there is no clear correlation between ``end of first plateau'' and pc 1. In contrast to linkwise migration, a larger mean degree is not correlated with more migration at the same value of $d$ for patchwise migration, and this explains the difference between the two migration modes. However, ``final position'' decreases with pc 1, although the correlation is weaker than for linkwise migration.
This means that for a larger mean degree the asymptotic robustness value is reached at a smaller migration strength $d$ even though a larger mean degree does not imply more migration. But a larger mean degree nevertheless allows for a faster equilibration of the population distribution over the patches, and this in turn is responsible for the asymptotic robustness value, as explained above. 

The relation between the value of the global maximum and the first pc seems to be very similar to the case of linkwise migration, as is to be expected if our explanation based on the rescue effect is correct.

When these four quantitites are plotted against pc 2, only "first 0.75" shows a clear trend, see  Figure \ref{pwpc2allinone}. This means that a large variation in degree between the different patches tends to drive species S$_2$ to extinction. In contrast to linkwise migration, patchwise migration causes a migration bias into highly connected patches, and this in turn produces new dynamical patterns as migration strength increases, making it apparently more likely that S$_2$ goes extinct. 

\begin{figure}[htp]
\centering
\includegraphics[width=10cm]{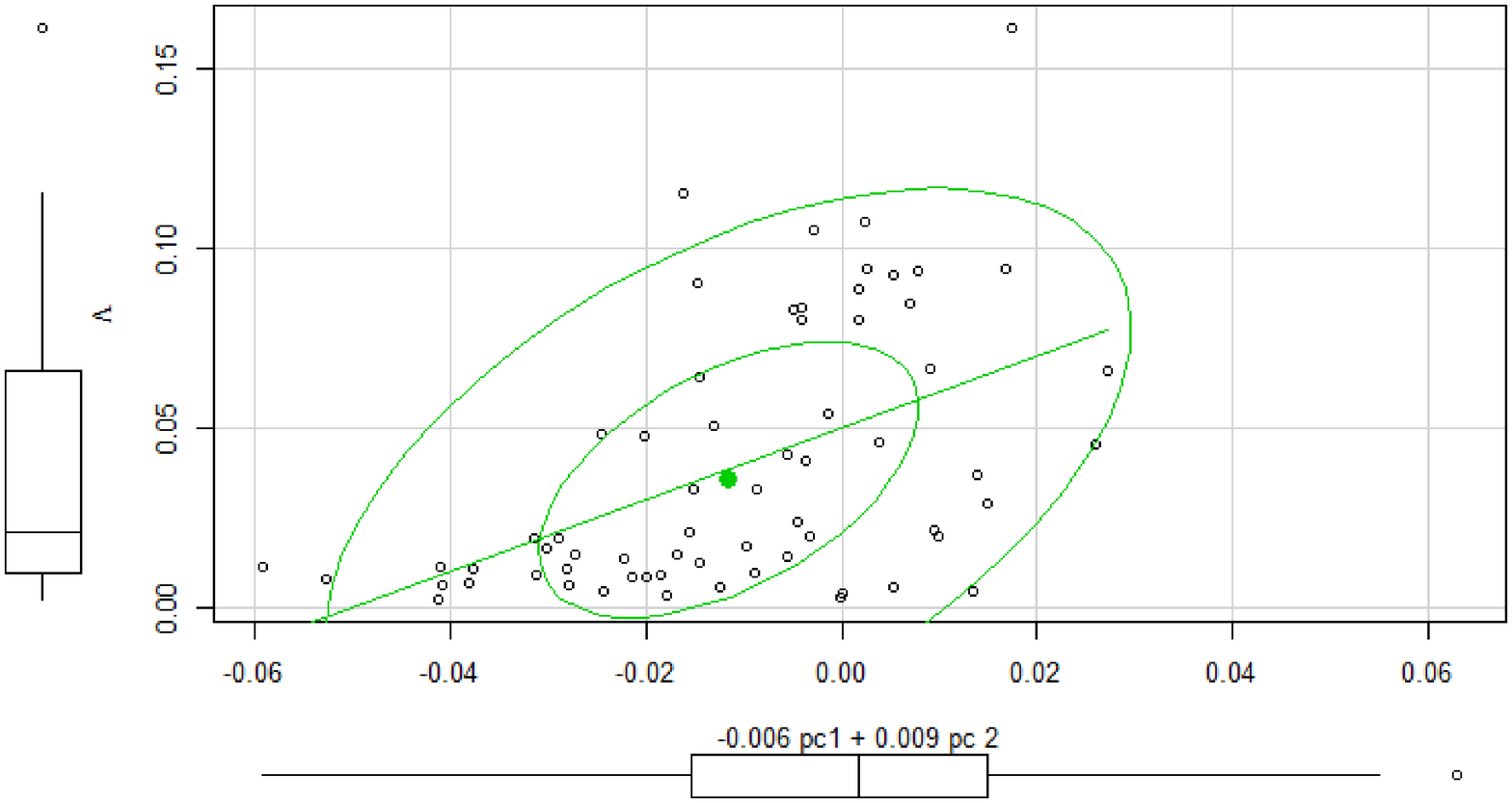}
\caption{The parameter $\Lambda$, which corresponds to the difference between the local maximum and minimum to the right of the first plateau, plotted against the linear combination $ -0.006 \text{ pc1} + 0.009 \text{ pc2} $. 
The ellipses contain 50 resp. 95 percent of all topologies, and the boxplots next to the axes indicate the distribution of the topologies with respect to the corresponding variable.}
\label{ellipse}
\end{figure}

For patchwise migration, we also evaluated the quantity $\Lambda$, which is the difference between the local maximum and minimum at the right-hand side of the first plateau and is defined only for those topologies that show this feature. We searched for the linear function of pc 1 and pc 2 that describes $\Lambda$ best. The result is $ -0.006 \text{ pc1} + 0.009 \text{ pc2} +0.05 $.  Figure \ref{ellipse} shows the plot. 

We also evaluated the linear combination $ -0.006 \text{ pc1} + 0.009 \text{ pc2} $ for those topologies that show a second plateau with a robustness value of 0.75 versus those that do not show it, and we found that the espective boxes limited by the upper and lower quantile do not overlap. 
 
The linear combination $-0.006 \text{ pc1} + 0.009 \text{ pc2}$ agrees well with the betweenness centrality, as can be concluded from Figure \ref{biplots}. Indeed, it is plausible that a topology with central patches through which a lot of migration flows leads to less stable dynamics.

\subsection{Comparison of linkwise and patchwise migration}
\begin{figure}[htp]
\centering
\includegraphics[width=10cm]{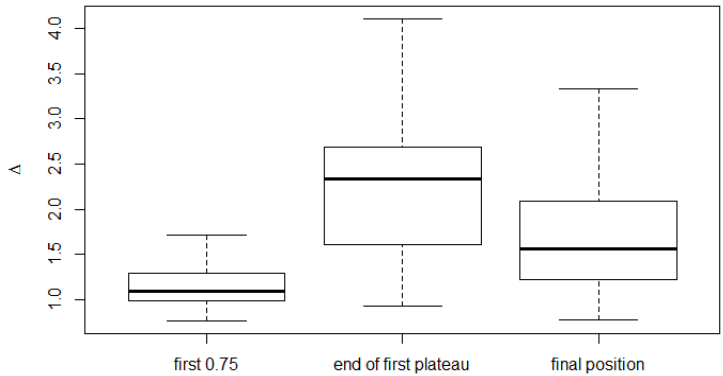}
\caption{Boxplot of the ratios $\Delta$ of ``first 0.75'', ``end of first plateau'' and ``final position''. The medians are indicated by the bold line and are 1.1, 2.3 and 1.6.   }
\label{ratioboxplots}
\end{figure}

We have seen that the position on the migration axis of the plateau with a robustness value of 0.75 decreases with the mean degree for linkwise migration, but not for patchwise migration. In order to verify our explanation that this is essentially due to the fact that a larger degree means more migration for linkwise migration, we evaluated the quantity $\Delta(x,T)$ given in equation (\ref{Delta}) for all three $x$. The result is shown in Figure \ref{ratioboxplots}. Indeed, $\Delta$ is centered aroud 1 for "first 0.75", which means that the difference in this quantity between linkwise and patchwise migration vanishes when the migration strength is corrected by the mean degree. However, this does not apply to ``end of first plateau'' and ``final position''. Here, the fact that patchwise migration causes an uneven distribution of populations over the patches appears to induce additional effects. 

For all topologies apart from the star topology, $T=0$, the global maximum of the robustness curves is essentially equal for patchwise and linkwise migration. For the star topology with patchwise migration, the difference in population sizes between different patches becomes most pronounced, and the central patch can act as a reservoir for all species, protecting them from extinction and thus causing a large robustness maximum \cite{plitzko2}.

\section{Conclusion}
\label{conclusion}

We have studied the influence of the topology of a six-patch migration network on the dynamics of the diamond food web module. The values of the parameters were chosen such that a single patch has a periodically oscillating attractor where all four species are present, and a fixed point attractor, where one species (labeled S$_2$) dies out and the other three species form a food chain. This system allows to evaluate robustness, which is the average number of species that survive population dynamics when starting from random initial conditions. Robustness is a measure of foodweb stability that is often evaluated when investigating larger food webs \cite{brose,heckmann}. Depending on the migration rate, the robustness varies between 0.75 and 1 in a non-monotonous way. This rich structure of the robustness curve makes this system particularly suitable to evaluate the correlation between dynamical behaviour and spatial topology. We chose a system of 6 patches because this provides a sufficiently large number of topologies (112) in order to perform a statistical evaluation. 

We found that a characterization of the topologies in terms of the mean degree and its standard deviation was sufficient for establishing the correlations between the dynamical and topological features. This is due to the fact that for the six-patch system other measures from graph theory are strongly correlated with these two measures. This holds even for measures that depend on the weights of connections, such as path lengths. 

When migration occurs through every connection with the same rate (``linkwise migration'') larger mean degree means more immigration into each patch and more emigration from each patch, given the same value of the migration strength $d$. As $d$ is increased, larger mean degree correlates therefore with an earlier onset of the dynamical extinction of species S$_2$ and an earlier arrival at the asymptotic (large-$d$) robustness value. When migration out of every patch occurs with the same rate (``patchwise migration''), the robustness curve still arrives earlier at the asymptotic value when the mean degree is larger, due to the easier equilibration of populations in the presence of more connections. The onset of the dynamical extinction of species S$_2$ correlates with the standard deviation of the degree distribution for patchwise migration, which seems plausible since a broader distribution of degrees allows for dynamics that are more different from that of an isolated patch. For patchwise migration, part of the topologies show a second plateau where the robustness value drops to 0.75. From recent work by Ristl et al. \cite{plitzko} we know that this happens at values of the migration rate where the oscillations on the different patches are synchronized to a large extent. We found that the existence of this second plateau, and more generally the depth of the dip on the right-hand side of the first plateau, correlates with the betweenness centrality of the topology. It appears that topologies where migration must go through a central patch are more prone to species extinction upon onset of synchronization. 

We also found that the star topology with patchwise migration leads to the largest robustness maximum. This is a feature that is also observed in systems with many more species \cite{plitzko2}. In general, the robustness curves become less rugged when the number of species is increased,  and the dominant trend for most topologies is an increase of robustness with migration rate \cite{plitzko2}. This is due to an averaging effect of the different species. The star differs from the other topologies since it has a very uneven distribution of the connections over the nodes. Since stars act as hubs on larger migration networks, the robustness peak on stars means that hubs can be important reservoirs for populations that are prone to extinction in less connected patches. The importance of hub nodes for the dynamics of networks appears to be a general feature, which is also seen in systems that are completely different \cite{muller2008organization}.

Our result that synchronization destabilizes the system  when the betweenness centrality of the migration network is larger is different in spirit from other studies on the relation between synchronization and instability. Since our model is constructed such that it has a stable oscillation on an isolated patch, a synchronized oscillation on coupled patches cannot unstable as long as the dynamics on every patch remains similar to that of an isolated patch. A reduction in robustness to 0.75 is in our model therefore due to the dynamical instability of the four-species oscillation in the coupled system. In contrast, other studies focus on the fact that synchronous oscillations in themselves can be considered as destabilizing, and they define a stability measure based on the summed species variability \cite{loreau2013}. This is because the populations of all patches go simultaneously through their minimum and can simultaneously become extinct due to a random perturbation. However, with asynchronous oscillations, a population that went extinct in one patch can be rescued by immigration from other patches where this population had a larger size at the moment of the random perturbation.

The study most closely related to ours is probably that by Moore at al \cite{horsthemke2004network}. These authors investigated a reaction-diffusion system (which is a two-species system) on all different topologies of 2,3, and 4 patches and established the conditions for the Turing instability, where a transition from a homogeneous fixed point to stationary patterns occurs. They find that linear arrays of patches are most susceptible to the Turing instability, while fully connected ones are most robust. These results bear some resemblance to our finding that a larger betweenness centrality is correlated with a smaller robustness when the migration strength $d$ is of the order of 1. 

Our findings do not depend on the precise number of patches. In fact, we investigated also the case of 5 patches and found similar results. When the number of patches becomes considerably larger, there exist topologies that differ more widely by their different measures, and this will certainly lead to interesting correlations with the dynamics. However, if this shall not be an abstract mathematical exercise, one has to choose realistic topologies for ecological systems, where patches are embedded in two-dimensional space and connections are placed in dependence of the geometric distance between patches \cite{urban2001landscape}. Interesting extensions of such approaches are obtained when the different dispersal abilities of different species are taken into account, such that each of them has a different migration network, leading to even more complex dynamics \cite{kouvaris2014pattern}.

To conclude, our study has demonstrated that a few topological measures of the migration network, together with the migration rate, are sufficient to predict the stability of a food-web module. These findings have relevance far beyond the field of ecology. There are many other examples of networks that live on several patches in space that are coupled by some type of transport or diffusion, such as signalling networks in tissues, chemical reaction networks in space, interacting traffic networks, and social networks. While the relation between topology and dynamics on simple networks is well understood for many systems, similar investigations on such complex multilayer networks have only recently begun. Our results demonstrate that the analysis of correlations between dynamical features and key topological measures can give deep insights into multilayer networks.

\section*{Funding} 
This work was supported by the Deutsche Forschungsgemeinschaft (Br2315/16-1, Dr300/12-1) within the  Research Unit FOR 1748.

\newpage
\newpage
\bibliographystyle{mystyle} 
\bibliography{lib}

\end{document}